\documentclass[prd,aps,showpacs,nofootinbib,onecolumn,superscriptaddress,
amssymb]{revtex4}

\usepackage{graphicx}
\usepackage[english]{babel}
\usepackage{amsmath}
\usepackage{amssymb}
\usepackage{amsfonts}
\usepackage{latexsym}

\newcommand{\be}{\begin{equation}}
\newcommand{\ee}{\end{equation}}
\newcommand{\bea}{\begin{eqnarray}}
\newcommand{\eea}{\end{eqnarray}}
\newcommand{\beaa}{\begin{eqnarray*}}
\newcommand{\eeaa}{\end{eqnarray*}}

\def\be{\begin{equation}}
\def\ee{\end{equation}}
\def\bea{\begin{eqnarray}}
\def\eea{\end{eqnarray}}

\begin{document}

\title{On the Expanding Phase of a Singular Bounce and Intermediate Inflation: The Modified Gravity Description}

\author{V.K. Oikonomou}
\email{v.k.oikonomou1979@gmail.com}
\affiliation{Tomsk State Pedagogical University, 634061 Tomsk, Russia}
\affiliation{Laboratory for Theoretical Cosmology, Tomsk State University of Control
Systems
and Radioelectronics (TUSUR), 634050 Tomsk, Russia}

\begin{abstract}
We demonstrate that the intermediate inflation scenario, is a
singular inflation cosmology, with the singularity at the origin
$t=0$ being a pressure and energy density singularity and
particularly a Type III singularity. Also, we show that the
expanding phase of a singular bounce, can be identical to the
intermediate inflation scenario, if the singular bounce has a Type
III singularity at the origin. For the intermediate inflation
scenario we examine the cosmological implications on the power
spectrum in the context of various forms of modified gravity.
Particularly we calculate the power spectrum in the context of
$F(R)$, $F(G)$ Gauss-Bonnet gravity and also for $F(T)$ gravity and
we discuss the viability of each scenario by comparing the resulting
spectral index with the latest observational data.
\end{abstract}


\maketitle

\section{Introduction}

Finite-time singularities \cite{Nojiri:2005sx} are timelike
singularities which frequently occur in modified gravity cosmologies
\cite{Nojiri:2005sx,Bamba:2008ut,Bamba:2012vg,Setare:2012vs,Myrzakul:2013qka,Kleidis:2016vmd},
and there exist various types of singularities of this sort, with
the most phenomenologically severe being the Big Rip. The most
phenomenologically ``soft'' type of singularity is the Type IV
\cite{Odintsov:2016plw,Oikonomou:2015qfh,Nojiri:2015qyc,Odintsov:2015gba,Odintsov:2015jca},
and the terminology ``soft'' refers to the fact that the Universe
can smoothly pass through these singularities and the physical
quantities are finite on these singularities. Type IV singularities
only cause dynamical instabilities and these may cause exit from
inflation in some inflationary models, see for example
\cite{Odintsov:2015gba,Odintsov:2015jca}. Also in the context of
bouncing cosmologies, it was demonstrated that it is possible for a
Type IV singularity to occur at the bouncing point, so a singular
bounce may occur \cite{Odintsov:2015ynk,Oikonomou:2015qha}. This
singular bounce however is called singular, but the singularity is
not of crushing type, and the effects of the singularity is
dynamical instabilities, see Ref.
\cite{Odintsov:2015gba,Odintsov:2015jca} for details.

On the other hand, it is possible that various inflationary
scenarios may contain some sort of singularity during the
inflationary era, or even at the end of the inflationary era
\cite{Odintsov:2015gba,Odintsov:2015jca}. One quite interesting
inflationary scenario is the intermediate inflation scenario
\cite{Barrow:1990td,Barrow:1993zq,Rezazadeh:2014fwa,Barrow:2006dh,Barrow:2014fsa,Herrera:2014mca,Jamil:2013nca,Herrera:2010vv,Rendall:2005if},
which in most cases it was realized in the context of scalar-tensor
cosmology. The aim of this paper is three-fold: Firstly we will
simply highlight the fact that the intermediate inflation scenario
is a singular inflationary cosmology, since it contains a Type III,
or pressure-energy density singularity at the origin $t=0$.
Secondly, we will demonstrate that the intermediate inflation
scenario can be identical with the expanding phase of a singular
bounce, with a Type III singularity occurring at the bouncing point.
Thirdly, we will examine whether the Type III singular bounce, or
equivalently the intermediate inflation scenario, can produce a
nearly scale invariant power spectrum compatible with the
observations, in the context of modified gravity. With regards to
modified gravity, we shall be interested for the $F(R)$ gravity, the
$F(G)$ gravity and the $F(T)$ gravity descriptions. The calculation
of the power spectrum is particularly easy, since these where
performed for the Type IV singular bounce
\cite{Odintsov:2015ynk,Oikonomou:2015qha}, so the case at hand is a
simple generalization.

This paper is organized as follows: In section II we briefly present
some essential features of finite-time singularities, and in
addition we demonstrate that the intermediate inflation scenario is
a Type III singular inflation scenario, according to the
classification of finite-time singularities. In addition, in section
II, we show that the expanding phase of a Type III singular bounce
can be identical with the intermediate inflation scenario, at least
functionally. In section III, we investigate when the power spectrum
of the intermediate inflation scenario in the context of $F(R)$,
$F(G)$ and $F(T)$ gravity, can be compatible with the observational
data. Finally, the conclusions follow in the end of the paper.

\section{Intermediate Inflation as Type II Singular Inflation and the Singular Bounce}

Before we start discussing the singularity structure of the
intermediate inflation scenario, we shall present some essential
information with regards to finite-time singularities. These were
firstly classified in Ref. \cite{Nojiri:2005sx}, from which we shall
use the terminology and notation. According to the classification in
\cite{Nojiri:2005sx}, there are four types of finite-time
singularity, varying from crushing types to softer types, as
follows:
\begin{itemize}
\item Type I Singularity (``The so-called Big Rip Singularity''): This is a crushing type singularity,
and therefore all the physical quantities defined on a three
dimensional spacelike hypersurface, which is determined by the time
instance that the singularity occurs, are singular. Particularly,
this is a timelike singularity which if it occurs at the time
instance $t=t_s$, then as $t\rightarrow t_s$, the scale factor
$a(t)$, the total energy density $\rho_{\mathrm{eff}}$ and the total
pressure $p_\mathrm{eff}$, strongly diverge, that is, $a \to
\infty$, $\rho_\mathrm{eff} \to \infty$ and
$\left|p_\mathrm{eff}\right| \to \infty$.
\item Type II Singularity (``The so-called Sudden Singularity''):
This is a pressure singularity, which firstly appeared in Refs.
\cite{Barrow:2004he,Barrow:2004hk,
Barrow:2004xh,FernandezJambrina:2004yy,Dabrowski:2004bz,Lake:2004fu,
Nojiri:2004ip,deHaro:2012wv}, and as it is obvious from the
terminology pressure singularity, only the pressure diverges as
$t\to t_s$, that is $\left|p_\mathrm{eff}\right| \to \infty$, but
both the scale factor and the total energy density are finite as
$t\to t_s$, that is $a \to a_s$, $\rho_{\mathrm{eff}}\to \rho_s$.
\item Type III Singularity: This type of singularity is the second
most sever after the Big Rip singularity, and in this case both the
pressure and the energy density diverge as $t\to t_s$, that is
$\left|p_\mathrm{eff}\right| \to \infty$ and $\rho_\mathrm{eff} \to
\infty$, but the scale factor is finite, $a \to a_s$. We shall call
this type of singularity the ``energy-pressure'' singularity.
\item Type IV Singularity: This singularity is the most ``harmless''
singularity from a phenomenological point of view, since all the
physical quantities are finite at $t\to t_s$, that is, $a \to a_s$,
$\rho_\mathrm{eff} \to \rho_s$ and $\left|p_\mathrm{eff}\right| \to
p_s$. However, in this case the higher derivatives $n\geq 2$ of the
Hubble rate diverge as $t\to t_s$. These singularities cause
dynamical instabilities on the inflationary dynamics
\cite{Odintsov:2015gba,Odintsov:2015jca} and also to cosmological
phenomenology at early-times, and these were extensively studied
further in \cite{Barrow:2015ora,Nojiri:2015fra}.
\end{itemize}
Having the classification for the finite-time singularities, we now
demonstrate that the intermediate inflation scenario has a Type III
singularity at the origin $t=0$. The intermediate inflation scale
factor is \cite{Barrow:1990td,Barrow:1993zq}, ,
\begin{equation}\label{bambabounce}
a(t)=e^{A\,t^n}\, ,
\end{equation}
with $0<n<1$ and also $A>0$, and the corresponding Hubble rate is,
\begin{equation}\label{hublerateinterm}
H(t)=A n t^{n-1}\, .
\end{equation}
The values of $n$ determine the singularity structure of the
intermediate inflation scenario, and specifically the singularity
structure depending on the value of $n$ is as follows,
\begin{itemize}\label{lista}
\item When $n<0$, a Type I singularity (Big-Rip) occurs at $t=0$.
\item When $0<n<1$, a Type III singularity occurs at $t=0$.
\item When $1<n<2$, a Type II singularity occurs at $t=0$.
\item When $n>1$, a Type IV singularity occurs at $t=0$.
\end{itemize}
Since for the intermediate inflation scenario, $n$ must be chosen as
$0<n<1$, then, from the classification above it is obvious that at
$t=0$ the intermediate inflation has a Type III singularity, which
is an energy-pressure singularity.

The singular bounce cosmology
\cite{Odintsov:2015ynk,Oikonomou:2015qha} has the same functional
form for the scale factor and for the Hubble rate as in the
intermediate inflation scenario, with the difference that the
parameter $n$ is restricted for physical consistency reasons as
follows,
\begin{equation}\label{rev}
n-1=-\frac{2n+1}{2m+1}\, .
\end{equation}
The restricted form for $n$ in Eq. (\ref{rev}) is necessary since we
need the scale factor to be real for $t<0$ and for $t>0$, and also
it is required that the Hubble rate is negative for $t<0$ and
positive for $t>0$. Hence if $n$ is chosen as in Eq. (\ref{rev}),
the Hubble rate for $t<0$ is negative and for $t>0$ it is positive,
which is the normal behavior expected for a bounce\footnote{Note
that for the number $(-1)^{1/3}$ has actually two complex branches
and only one real negative, and we choose the negative branch.}. In
effect, if $n$ is chosen as in Eq. (\ref{rev}), and also if $0<n<1$,
the bounce occurs, however it has a Type III singularity at the
origin. Effectively, it differs from other bouncing cosmologies,
since it has a singularity at the bouncing point $t=0$, however our
aim was to show that the expanding phase of this Type III singular
bounce and of the intermediate inflation scenario are identical, at
least functionally. It is conceivable however, that the primordial
perturbations in the two scenarios will have different origin.

\section{Intermediate Inflation from Modified Gravity and Comparison with Observations}

The functional similarity of the intermediate inflation scenario
with the singular bounce will enable us to provide analytic results
on the power spectrum of primordial curvature perturbations, and the
corresponding spectral index, by using the related literature. For
example, the $F(R)$ gravity description of the singular bounce was
performed in Ref. \cite{Odintsov:2015ynk}, and the $F(G)$
description of the singular bounce was performed in
\cite{Oikonomou:2015qha}, so in these sections we share the results
of these works and we shall apply these for the case of the
intermediate inflation scenario.

We start off with the $F(R)$ gravity description, in vacuum, with
the $F(R)$ gravity action being,
\begin{equation}
\label{action1dse} \mathcal{S}=\frac{1}{2\kappa^2}\int
\mathrm{d}^4x\sqrt{-g}F(R)\, ,
\end{equation}
and as it was shown in Ref. \cite{Odintsov:2015ynk}, the $F(R)$
gravity which realizes the cosmology (\ref{bambabounce}) during the
early time era, has the following form,
\begin{equation}\label{jordanframegravity}
F(R)=R+a_2R^2+a_1 \, ,
\end{equation}
with the parameters $a_1$ and $a_2$ being positive and can be found
in the Appendix of Ref. \cite{Odintsov:2015ynk}. Following
\cite{Odintsov:2015ynk}, the corresponding power spectrum
$\mathcal{P}_R$ at early times has the following form,
\begin{equation}\label{powerspectrumfinal}
\mathcal{P}_R\sim k^{3+\frac{n-4\mu-2\mu (n-1)}{n-1}}\, .
\end{equation}
where the parameter $\mu$ is $\mu=\frac{1}{2} (2n-1) )$. Obviously,
the power spectrum is not scale invariant, but now we investigate
whether the resulting spectral index can be compatible with the 2015
Planck data \cite{Ade:2015lrj}, for some value in the range $0<n<1$,
which corresponds to the intermediate inflation scenario. The
spectral index of the power spectrum $\mathcal{P}_R$, is defined as
follows,
\begin{equation}\label{fieldns}
 n_s-1\equiv\frac{d\ln\mathcal{P}_{\mathcal{R}}}{d\ln
k},
\end{equation}
and we easily find that, the spectral index of the power spectrum of
Eq. (\ref{powerspectrumfinal}), is equal to,
\begin{equation}\label{ns}
n_s=4+\frac{n-4\mu-2\mu (n-1)}{n-1}\, .
\end{equation}
The 2015 Planck data \cite{Ade:2015lrj}, constraint the spectral
index as follows,
\begin{equation}
\label{planckdata} n_s=0.9644\pm 0.0049\, ,
\end{equation}
and it can be shown that the spectral index of Eq. (\ref{ns}) cannot
be compatible with the constraint (\ref{planckdata}), for any value
of $n$ in the range $0<n<1$.

Now we turn our focus on the modified Gauss-Bonnet case, in which
case, the vacuum $F(G)$ gravity action has the following form,
\begin{equation}\label{actionfggeneral}
\mathcal{S}=\frac{1}{2\kappa^2}\int \mathrm{d}^4x\sqrt{-g}\left (
R+F(G)\right )\, ,
\end{equation}
with $\kappa^2=1/M_{pl}^2$, and also $M_{pl}=1.22\times 10^{19}$GeV.
As it was shown in Ref. \cite{Oikonomou:2015qha}, the $F(G)$ gravity
which realizes the singular bounce and in effect the intermediate
inflation for $0<n<1$, has the following form at early times,
\begin{equation}\label{fgsmalglimit}
F(G)\simeq C_2 G+B G^{\frac{\alpha }{-1+3 \alpha }}\, .
\end{equation}
The resulting spectrum was found to be as follows,
\begin{equation}\label{powerspectrumfinal}
\mathcal{P}_R\sim k^{\frac{7}{2}+3+\frac{\left(2-2 (n-1) +(n-1)
^2\right) \mu }{2 (n )}}\, ,
\end{equation}
with $\mu=-11/n$. Accordingly, the resulting spectral index is equal
to,
\begin{equation}
 n_s-1\equiv\frac{d\ln\mathcal{P}_{\mathcal{R}}}{d\ln
k}\simeq 1-\frac{11}{2n^2}\, .
\end{equation}
Then, it can be easily shown that the compatibility with Planck data
\cite{Ade:2015lrj} comes when $12.9<n<13.7$, and therefore the
intermediate inflation inflation scenario in the context of modified
Gauss-Bonnet gravity, is not compatible with the observational data.

The picture is entirely different when the intermediate inflation
scenario is realized in the context of $F(T)$ gravity, see Ref.
\cite{Cai:2015emx} for a review. We shall briefly present the
calculation of the primordial curvature perturbations, for the
intermediate inflation scenario in the context of $F(T)$ gravity.
The details of this calculation will be presented elsewhere
\cite{ftsub}. Studies on the evolution of primordial perturbations
in $F(T)$ gravity were performed in Refs.
\cite{Cai:2011tc,Chen:2010va,Izumi:2012qj,Nashed:2014lva,Hanafy:2014bsa,Hanafy:2014ica},
and we follow \cite{Cai:2011tc}. The perturbed metric in the
longitudinal gauge is,
\begin{equation}\label{metricscalar}
\mathrm{d}s^2=(1+2\Phi)\mathrm{d}t^2-a(t)^2(1-2\Psi)\sum_i\mathrm{d}x^2_i\,
,
\end{equation}
so the scalar functions $\Phi$ and $\Psi$ practically quantify the
scalar fluctuations of the metric. We can express the torsion scalar
as a function of  $\Phi$ and $\Psi$ in the following way,
\begin{equation}\label{trosionpertubrscalar}
\delta T=12 H(\dot{\Phi}+H\Psi)\, ,
\end{equation}
with $H$ being the Hubble rate. By assuming an $F(T)$ gravity of the
form $F(T)=T+f(T)$, the perturbation equations in $F(T)$ gravity are
\cite{Cai:2011tc},
\begin{align}\label{ftgravieqnspertrb}
&
(1+f_{,T})\frac{\nabla^2}{a^2}\Psi-3(1+f_{,T})H\dot{\Psi}-3(1+f_{,T})H^2\Phi\\
\notag & +36f_{,TT}H^3(\dot{\Psi}+H\Phi)=4\pi G \delta \rho\, ,\\
\notag &
(1+f_{,T}-12H^2f_{,TT})(\dot{\Psi}+H\Phi)=4\pi G\delta q\, ,\\
\notag &(1+f_{,T})(\Psi-\Phi)=8\pi G\delta s\, , \\ \notag &
(1+f_{,T}-12H^2f_{,TT})\ddot{\Psi}+3H(1+f_{,T}
\\
\notag & -12H^2f_{,TT}-12\dot{H}f_{,TT}+48H^2\dot{H}f_{,TTT})\dot{\Psi}\\
\notag & +\Big{[}
3H^2(1+f_{,T}-12H^2f_{,TT})+2\dot{H}(1+f_{,T}-30H^2f_{,TT}\\
\notag &
+72H^4f_{,TTT})\Big{]}\Phi+\frac{1+f_{,T}}{2a^2}\nabla^2(\Psi-\Phi)=4\pi
G \delta p \, ,
\end{align}
where $f_{,T}$, stands for $\partial_T f(T)$, and the derivatives
$f_{,TT}$ and $f_{,TTT}$ are defined accordingly. Also the functions
$\delta p$, $\delta \rho$, $\delta s$, $\delta q$, stand for the
fluctuations of the total pressure, of the total energy density, of
the anisotropic stress and of the fluid velocity respectively. If
the matter fluids present are represented by a canonical scalar
field with potential $V(\phi)$, we obtain the following equations,
\begin{align}\label{deltarhodrelations}
& \delta
\rho=\dot{\phi}(\delta\dot{\phi}-\dot{\phi}\Phi)+V_{,\phi}\delta
\phi\, ,\\ \notag & \delta q=\dot{\phi}\delta \phi\, ,\\ \notag &
\delta s=0\, ,\\ \notag & \delta p=\dot{\phi}(\delta
\dot{\phi}-\dot{\phi}\Phi)-V_{,\phi}\delta \phi\, .
\end{align}
As was shown in \cite{Cai:2011tc}, due to the above relations we get
$\Psi=\Phi$, and therefore, the scalar fluctuation $\delta \phi$
determines the gravitational potential $\Phi$, and thus we have a
single degree of freedom. In effect, the evolution of scalar
perturbations is determined by the following differential equation
\cite{Cai:2011tc},
\begin{equation}\label{masterequation}
\ddot{\Phi}_k+\alpha
\dot{\Phi}_k+\mu^2\Phi_k+c_s^2\frac{k^2}{a^2}\Phi_k=0\, ,
\end{equation}
where $\Phi_k$ is the scalar Fourier mode of the potential $\Phi$,
and moreover, the functions $\alpha$, $c_s^2$ and $\mu^2$ are the
the frictional term, the speed of sound parameter and the effective
mass respectively,corresponding to the scalar potential $\Phi$.
These functions are defined as follows,
\begin{align}\label{functionsanalytical}
&
\alpha=7H+\frac{2V_{,\phi}}{\dot{\phi}}-\frac{36H\dot{H}(f_{,TT}-4H^2f_{,TTT})}{1+f_{,T}-12H^2f_{,TT}}\,
,\\ \notag &
\mu^2=6H^2+2\dot{H}+\frac{2HV_{,\phi}}{\dot{\phi}}-\frac{36H^2\dot{H}(f_{,TT}-4H^2f_{,TTT})}{1+f_{,T}-12H^2f_{,TT}}\,
, \\ \notag & c_s^2=\frac{1+f_{,T}}{1+f_{,T}-12H^2f_{,TT}}\, .
\end{align}
The equation of motion for the canonical scalar field is,
\begin{equation}\label{eqnforscaux}
\ddot{\phi}+3H\dot{\phi}+V_{,\phi}=0\, ,
\end{equation}
and therefore the $f(T)$ gravity equations of motion can be written
as follows,
\begin{equation}\label{secondfriedmann}
(a+f_{,T}-12H^2f_{,TT})\dot{H}=-4\pi G \dot{\phi}^2\, .
\end{equation}
In effect, the equation that determines the evolution of the scalar
perturbations is,
\begin{equation}\label{evolutionequationfinal}
\ddot{\Phi}_k+\left(H-\frac{\ddot{H}}{\dot{H}}\right)\dot{\Phi}_k+\left(2\dot{H}-\frac{H\ddot{H}}{\dot{H}}
\right)\Phi_k+\frac{c_s^2k^2}{a^2}\Phi_k=0\, .
\end{equation}

A gauge invariant physical quantity that quantifies any cosmological
inhomogeneities, is the comoving curvature fluctuation, denoted as
$\zeta$, which is,
\begin{equation}\label{comovcurv}
\zeta =\Phi-\frac{H}{\dot{H}}\left (\dot{\Phi}+H\Phi \right)\, .
\end{equation}
We introduce the quantity $v$ defined as follows,
\begin{equation}\label{varintr1}
v=z\zeta\, ,
\end{equation}
with $z$ standing for,
\begin{equation}\label{fgr}
z=a\sqrt{2\epsilon}\, ,
\end{equation}
and also $\epsilon$ is $\epsilon=-\frac{\dot{H}}{H^2}$. Then the
equation that determines the evolution of perturbations is
\cite{Cai:2011tc},
\begin{equation}\label{mastereqn2}
v_k''+\left( c_s^2k^2-\frac{z''}{z}\right) v_k=0\, ,
\end{equation}
with $c_s$ appearing in Eq. (\ref{functionsanalytical}). The
``prime'' above indicates differentiation with respect to the
conformal time, defined as follows,
\begin{equation}\label{conformtime}
\tau=\int \mathrm{d}t\frac{1}{a}\, .
\end{equation}
The first Friedmann equation in $f(T)$ gravity is,
\begin{equation}\label{fteqn1}
H^2=-\frac{f(T(t))}{6}-2f_{,T}H^2\, ,
\end{equation}
and due to the fact that $T=-6H^2$, for the intermediate inflation
case we have,
\begin{equation}\label{explicitttrelation}
T=-6 A^2 n^2 t^{2 n-2}\, .
\end{equation}
By inverting the above and using Eq. (\ref{fteqn1}), we easily
obtain the approximate $f(T)$ gravity realizing the intermediate
inflation scenario, which is,
\begin{equation}\label{ftgravitytyfn1}
f(T)=c_1 T^{\frac{A n}{2}}-\frac{T}{2 \left(1-\frac{A
n}{2}\right)}\, ,
\end{equation}
where $c_1$ is an arbitrary integration constant. In effect, the
total $F(T)$ gravity is $F(T)=T+f(T)$. Due to the fact that we are
interested in early times, the exponential $e^{At^n}$ is
approximated as $e^{At^n}\sim 1$, and via the relation
(\ref{conformtime}) we can see that the conformal time can be
identified with the cosmic time. Therefore, we have for the
intermediate inflation scenario,
\begin{align}\label{zrt}
& z(t)=\sqrt{\frac{2 (1-n) t^{-n}}{A n}}\, , \\ \notag &
 c_s^2(t)=\frac{A \,c_1\, n 6^{\frac{A n}{2}} (A n-2) \left(-A^2
n^2 t^{2 n-2}\right)^{\frac{A n}{2}}}{2 (A n-1) S(t) } -\frac{12 A^2
n^2 (A n-1) t^{2 n-2}}{2 (A n-1) S(t)} \, ,
\end{align}
with $S(t)$ being,
\begin{align}\label{akyroeqns}
& S(t)=\Big{(}c_1 6^{\frac{A n}{2}} (A n-2) \left(-A^2 n^2 t^{2
n-2}\right)^{\frac{A n}{2}} -6 A^2 n^2 t^{2 n-2}\Big{)}\, .
\end{align}
Hence, at leading order, the equation which gives the evolution of
perturbations becomes,
\begin{equation}\label{mastereqn23}
v''_k(t)+\left( k^2-\frac{\left(\frac{n}{2}+1\right) n}{2
t^2}\right) v_k(t)=0\, ,
\end{equation}
which can be solved to yield at leading order \cite{ftsub},
\begin{equation}\label{dominanthubbleev}
v_k(t)=\frac{C_2 \sqrt{t} \left(2^{\frac{n}{2}+\frac{1}{2}} \Gamma
\left(\frac{n+1}{2}\right) (k
t)^{-\frac{n}{2}-\frac{1}{2}}\right)}{\pi }\, .
\end{equation}
The power spectrum is defined as follows,
\begin{equation}\label{powerspectrmrelation}
\mathcal{P}_{\zeta}=\frac{k^3}{2\pi^2}|\frac{v_k}{z}|_{k=a H}^2\, ,
\end{equation}
so by using the previous relations, it can be shown that
\cite{ftsub},
\begin{equation}\label{powerspectrumfinalrelationforft}
\mathcal{P}_{\zeta}\simeq \frac{\left(n^{1-\frac{n}{n-1}}
A^{1-\frac{n}{n-1}}\right) k^{\frac{1}{n-1}+3}}{4 \pi ^2 (1-n)}\, .
\end{equation}
Therefore, the corresponding spectral index is,
\begin{equation}\label{nsfinalforft}
n_s=\frac{1}{n-1}+4\, .
\end{equation}
The 2015 Planck data indicate that the spectral index has to be in
the interval $n_s=[0.9595,0.9693]$, and it can be shown that in
order for this to occur, the parameter $n$ in Eq.
(\ref{nsfinalforft}) has to be chosen in the interval
$n=[0.67,0.6711]$. Hence compatibility with the Planck can be
achieved for the intermediate inflation scenario, in the context of
$F(T)$ gravity, by appropriately choosing the parameters of the
theory.

\section{Conclusions}

In this work we showed that the intermediate inflation scenario is a
singular type of inflation, with the singularity occurring at the
origin being a Type III singularity. The intermediate inflation
scenario can be identical with the expanding phase of a Type III
singular bounce cosmology. By using the existing results in the
literature, we calculated the power spectrum of primordial curvature
perturbations for the intermediate inflation scenario, in the
context of vacuum $F(R)$, vacuum $F(G)$ and vacuum $F(T)$ gravity.
As we demonstrated, only the $F(T)$ gravity realizations yielded
results compatible with the 2015 observational data.

Our results show an important feature of modified gravities in
general. Particularly, it might occur that a modified gravity
description might be viable or not, in the context of some theory,
so this indicates that not all modified gravity descriptions yield
equivalent results. To our opinion, the theory which is actually
closer to physical reality will have the characteristic of being
compatible with most of the observations, and also it will be at the
same time, relatively simple, but ingeniously constructed. Here we
aimed to show that many different descriptions of modified gravity
are not equivalent.

\section*{Acknowledgments}

This work is supported by Ministry of Education and Science of
Russia (V.K.O).

\end{document}